\documentclass[12pt]{article}
\usepackage{graphicx}
\usepackage{float}
\usepackage{subfig}
\usepackage{lineno}

\newcommand\pubnumber{SNSN-323-63}
\newcommand\pubdate{\today}

\def\institute{INFN Sezione di Bologna\\
Via Irnerio 46, 40126 Bologna, ITALY}
\def\support{\footnote{Copyright 2018 CERN for the benefit of the ATLAS Collaboration.
CC-BY-4.0 license}}

\def\Title#1{\begin{center} {\Large #1 } \end{center}}
\def\Author#1{\begin{center}{ \sc #1} \end{center}}
\def\Address#1{\begin{center}{ \it #1} \end{center}}

\newcommand\pubblock{\rightline{\begin{tabular}{l} \pubnumber\\
         \pubdate  \end{tabular}}}
\newenvironment{Abstract}{\begin{quotation}  }{\end{quotation}}
\newenvironment{Presented}{\begin{quotation} \begin{center} 
             PRESENTED AT\end{center}\bigskip 
      \begin{center}\begin{large}}{\end{large}\end{center} \end{quotation}}

\def\beq{\begin{equation}}
\def\eeq#1{\label{#1}\end{equation}}
\def\eeqn{\end{equation}}

\def\beqa{\begin{eqnarray}}
\def\eeqa#1{\label{#1}\end{eqnarray}}
\def\eeqan{\end{eqnarray}}

\let\bar=\overbar

\def\Dslash{\not{\hbox{\kern-4pt $D$}}}
\def\dslash{\not{\hbox{\kern-2pt $\del$}}}

\def\msb{{\bar{\ssstyle M \kern -1pt S}}}

\begin{document}
\begin{titlepage}
\pubblock

\vfill
\Title{Differential measurements of $t \overline{t}$ production in ATLAS}
\vfill
\Author{ Matteo Negrini\\on behalf of the ATLAS Collaboration\support}
\Address{\institute}
\vfill
\begin{Abstract}
Differential cross sections of $t \overline{t}$ production have been measured 
by the ATLAS experiment at the LHC. 
Monte Carlo calculations provide an overall good modeling of all measured 
distributions, except for the transverse momentum of the top quark, 
for which all the generators at the next-to-leading order produce harder spectra. 
These measurements have been also included in the PDF fits, contributing 
to a sensible reduction of the gluon PDF uncertainty at large $x$.
\end{Abstract}
\vfill
\begin{Presented}
$11^\mathrm{th}$ International Workshop on Top Quark Physics\\
Bad Neuenahr, Germany, September 16--21, 2018
\end{Presented}
\vfill
\end{titlepage}
\def\thefootnote{\fnsymbol{footnote}}
\setcounter{footnote}{0}

\section{Introduction}
We will discuss some recent measurements performed by the ATLAS experiment \cite{ATLAS}
at the CERN Large Hadron Collider (LHC) of top quark-antiquark pair ($t \overline{t}$)
differential cross sections, 
both absolute or normalized to the total cross section (``shape'' measurements).
The measurements of these cross sections allow 
to test QCD models, to possibly observe unexpected effects in the top quark sector, 
and to improve all searches for new physics phenomena at the LHC in which $t \overline{t}$ 
production represents one major background source.
Here we will present a study of the constraints on the gluon parton distribution functions (PDF) 
of the proton from the $t \overline{t}$ differential cross sections. 

The differential cross sections are corrected for efficiency and resolution effects 
by means of unfoding techniques. 
Unfolding corrections can be evaluated from Monte Carlo (MC) samples to obtain measurements 
at the parton-level or particle-level. 
Parton-level measurements refer to the MC top quark after radiation but before decay and 
are extrapolated to the full phase-space. Particle-level ones refer to 
reconstructed objects analogous to the one defined at the detector-level but 
using stable particles in the MC (with life-time $\tau > 0.3 \times 10^{-10}$ s) and are 
extrapolated in a fiducial region closely following the event selection.

Systematic uncertainties affecting the measurements are due to the imperfect detector 
modeling and calibration and to the uncertainties in the background predictions. 
The related uncertainties on the cross section measurements 
are estimated by repeating the full analysis after $1\sigma$ 
variation of each source of systematic uncertainty. Modeling uncertainties, 
affecting the unfolding corrections, are estimated as the difference between the unfolded 
and true MC sample distributions after variation of the modeling assumptions for the 
$t \overline{t}$ matrix element generator, the parton shower (PS) and hadronization, 
initial and final state radiation (IFSR), and PDF.

For each differential cross section, a covariance matrix ($Cov$) is obtained from 
all the uncertainties affecting the measurement, including bin-to-bin correlations.
Measurements are compared with MC calculations at the next-to-leading (NLO) order in QCD 
coupled with PS, by means of $\chi^2 = V^T \cdot Cov^{-1} \cdot V$, 
where $V$ is the vector of differences between the measurement and the prediction.

\section{Lepton+jets channel}
Several kinematic distributions of $t \overline{t}$ events, such as the transverse 
momentum ($p_{\rm T}$) and rapidity ($y$) of the top quark and the $t \overline{t}$ system, 
and the mass of the $t \overline{t}$ system, 
have been recently measured at the particle level in the lepton+jets channel 
using $pp$ collisions at $\sqrt{s}=13$ TeV \cite{ljet}.
The recostruction of the hadronically decaying top quark 
is done by applying both ``resolved'' techniques, 
combining three anti-$k_T$ jets with radius R=0.4, 
and ``boosted'' ones, for events in which the top is produced at 
high $p_{\rm T}$ and the decay products 
are contained in a single broader jet, with R=1.0.

Good modeling is observed for different MC generators, except for the top $p_{\rm T}$
whose distribution, in all MC generators, tends to be harder than the measured one, and 
for the invariant mass of the $t \overline{t}$ system obtained using Herwig++. 

\begin{figure}[htb]
\centering
\subfloat[]{\includegraphics[height=0.4\textheight]{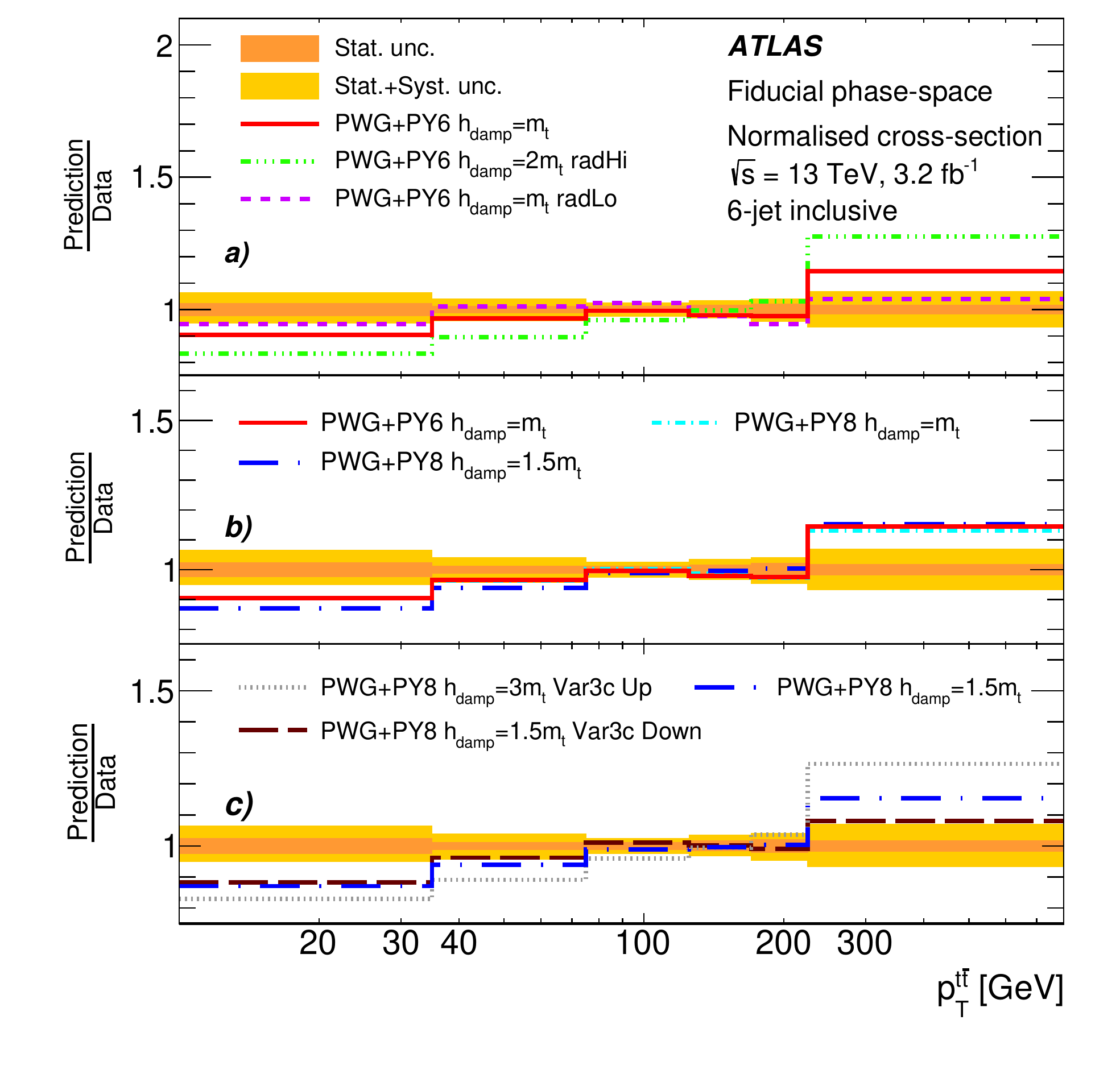}}
\hspace{0.0cm}
\subfloat[]{\includegraphics[height=0.4\textheight]{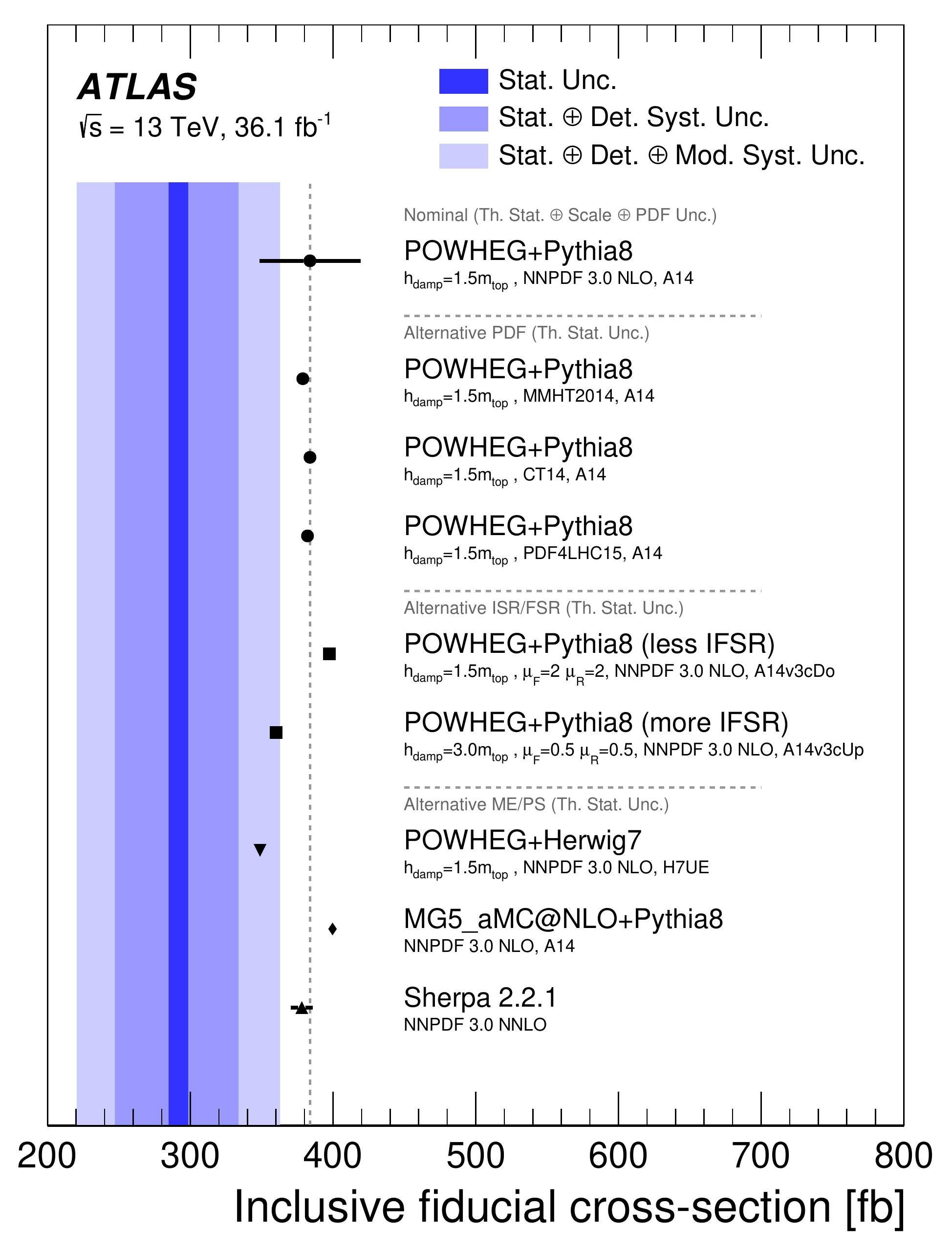}}
\caption{(a) Ratios of normalised differential cross sections as a 
function of $p_{\rm T}^{t \overline{t}}$ for different MC models with respect to data, 
for events with at least 6 jets. Figure from \cite{ljet+jets}.
(b) All-hadronic boosted fiducial cross section measurement at particle level 
(blue bands indicating the uncertainties) compared with Monte Carlo calculations. 
The uncertainty on the first model includes statistical, scale and PDF uncertainties. 
Figure from \cite{allhad}. }
\label{fig:diffxsec}
\end{figure}

A dedicated measurement of quantities particularly sensitive to 
the presence of gluon radiation in $t \overline{t}$ events, such as the $p_{\rm T}$ of the $t \overline{t}$ system ($p_{\rm T}^{t \overline{t}}$), 
the $p_{\rm T}$ of the hadronic top, and out-of-plane transverse momentum of the $t \overline{t}$ 
system ($p_{out}^{t \overline{t}}$) \footnote{$p_{out}^{t \overline{t}}$ is 
defined as the momentum of the hadronic top projected on the normal to the 
plane defined by the other top and the beam axis.} 
is done using a $t \overline{t}$ sample with extra jets \cite{ljet+jets}. 
The results are obtained for different number of extra jets and a good modeling of MC 
generators is generally observed, except for the $p_{\rm T}^{t \overline{t}}$ at large ($\geq 6$) 
mutiplicity of jets in the event, as shown in Fig. \ref{fig:diffxsec}(a). 

\section{All-hadronic boosted}
Differential cross sections for events with top quarks produced at 
high $p_{\rm T}$ are measured, at particle and parton levels, using a $\sqrt{s}=13$ TeV 
$pp$ collisions sample \cite{allhad}.
Top and antitop quarks are reconstructed in the all-hadronic channel 
by means of two large-R jets, and tagged as top quarks 
on the basis of jet substructure variables. This helps in reducing 
multi-jet QCD events that represent the main 
background affecting all-hadronic $t \overline{t}$ measurements. 

The multi-jet background is estimated by defining 15 control and validation regions, 
with variable but sub-dominant $t \overline{t}$ contribution, and one signal region. 
The region definition is based on the number of top-tagged large-R jets and 
$b$-tagged small-R sub-jets in the event.
The number of background events in the signal region is then obtained from the 
ratios of the events in the control regions. The purity 
of the selected sample in the signal region is 75\%.

Systematic uncertainties are typically at the few-percent level and are dominated by 
large-R jet calibration (JES calibration and modeling of quantities used in top tagging), 
but this analysis is focusing on a low-populated region of the $t \overline{t}$ production 
phase space so the statistical uncertainty is not negligible, especially for $p_{T,top}$ in
the TeV region. 
Also in this analysis, a good modeling of the reconstructed observables is observed 
for MC generators, except for an excess in the signal event yield in MC, consistent with 
previous observations that MC generators tend to show a harder $p_{\rm T}$ spectra 
than the measured one. 
The measured inclusive fiducial cross section at particle-level is 
$\sigma_{fid}=292 \pm 7~({\rm stat}) \pm 71~({\rm syst})$ fb  
and is lower than all predictions, as shown in Fig. \ref{fig:diffxsec}(b).

\section{Constraining the proton PDF using $t \overline{t}$ differential cross sections}
Top quark data have been added to a previous fit of deep inelastic scattering HERA data 
and ATLAS differential $W$ and $Z/\gamma^*$ cross sections \cite{ttbarPDF}. 
Differential measurements at $\sqrt{s}=8$ TeV of $t \overline{t}$ 
distributions that are more sensitive to the gluon PDF 
(such as the mass and rapidity of the $t \overline{t}$ system, 
and the average top quark $p_{\rm T}$), corrected at the parton-level 
exploiting NNLO theoretical calculations, are used.
Systematic uncertainties are fully correlated, among bins of the same distribution 
and between spectra. 

\begin{figure}[htb]
\centering
\includegraphics[width=0.6\textwidth]{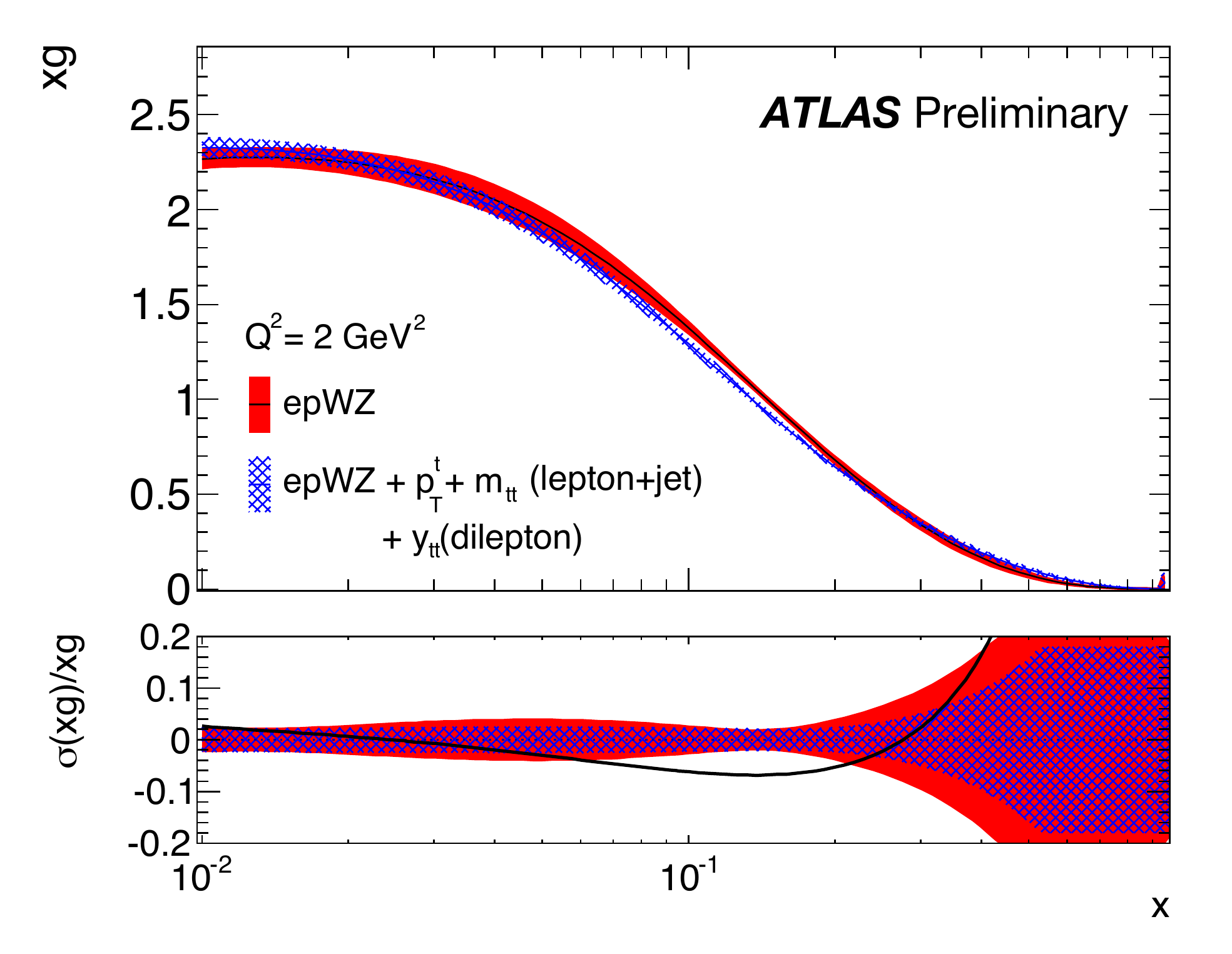}
\caption{The gluon spectrum from fitting HERA data and ATLAS $W,Z/\gamma^*$ boson data with
and without the ATLAS $t \overline{t}$ data. Figure from \cite{ttbarPDF}. }
\label{fig:pdf}
\end{figure}

Fig. \ref{fig:pdf} show that adding $t \overline{t}$ data produces 
a sensible reduction of the gluon PDF uncertainty in the large-$x$ region 
as well as a variation of the shape for $x>10^{-1}$.

\section{Summary}
Differential $t \overline{t}$ production cross sections, for several observables and 
in different reconstruction channels, have been measured by the ATLAS experiment at the LHC.
These measurements can be used to test QCD calculation and provide insights on the gluon PDF of the proton.
An overall good modeling of $t \overline{t}$ production is provided by the tested MC 
generators at the NLO, 
with the exception of the top quark $p_{\rm T}$ spectrum that tends to be harder in MC than in data. 
Possible ways to improve the measurements, in particular for high $p_{\rm T}$ top quarks, 
include a reduction of the large-R jet calibration and top-tagging uncertainties.

\end{document}